\documentclass[prd,twocolumn]{revtex4}%
\usepackage{amsfonts}
\usepackage{amsmath}
\usepackage{amssymb}
\usepackage{graphicx}%
\providecommand{\U}[1]{\protect\rule{.1in}{.1in}}
\begin{document}
\preprint{arXvir:0905.????}
\title[Bianchis \mathrm{VI}$_{0}\&\mathrm{III.}$]{Bianchi $\mathrm{VI}_{0}\&\mathrm{III}$ models. Self-similar approach. }
\author{Jos\'e Antonio Belinch\'on}
\email{abelcal@ciccp.es}
\affiliation{Dept. F\'{\i}sica. ETS Arqitectura. UPM. Av. Juan de Herrera 4. Madrid 28040. Spain}
\date{\today}

\pacs{PACS number}

\begin{abstract}
We study several cosmological models with Bianchi \textrm{VI}$_{0}%
\&\mathrm{III}$ symmetries under the self-similar approach. We find new
solutions for the \textquotedblleft classical\textquotedblright\ perfect fluid
model as well as for the vacuum model although they are really restrictive for
the equation of state. We also study a perfect fluid model with time-varying
constants, $G$ and $\Lambda.$ As in other models studied we find that the
behaviour of $G$ and $\Lambda$ are related. If $G$ behaves as a growing time
function then $\Lambda$ is a positive decreasing time function but if $G$ is
decreasing then $\Lambda_{0}$ is negative. We end by studying a massive cosmic
string model, putting special emphasis on calculating the numerical values of
the equations of state. We show that there is no SS solution for a string
model with time-varying constants.

\end{abstract}
\maketitle

\section{Introduction.}

The current observations of the large scale cosmic microwave background
suggest us that our physical universe is expanding, isotropic and homogeneous
models with a positive cosmological constant. The analysis of CMB fluctuations
could may confirm this picture. But other analysis reveal some inconsistency.
Analysis of WMAP data sets show us that the universe could have a preferred
direction. For this reason Bianchi models are important in the study of anisotropies.

The study of SS models is quite important since a large class of orthogonal
spatially homogeneous models are asymptotically self-similar at the initial
singularity and are approximated by exact perfect fluid or vacuum self-similar
power law models. Exact self-similar power-law models can also approximate
general Bianchi models at intermediate stages of their evolution. This last
point is of particular importance in relating Bianchi models to the real
Universe. At the same time, self-similar solutions can describe the behaviour
of Bianchi models at late times i.e. as $t\rightarrow\infty$ (see
\cite{ColeyDS})$.$Of particular interest is determining the exact value of the
equations of state (see \cite{Wainwrit}). The geometry and physics at
different points on an integral curve of a homothetic vector field (HVF)
differ only by a change in the overall length scale and in particular any
dimensionless scalar will be constant along the integral curves. In this sense
the existence of a HVF is a weaker condition than the existence of an KVF
since the geometry and physics are completely unchanged along the integral
curves of \ a Killing vector field (KVF). However, the existence of a
non-trivial HVF leads to restrictions on the equations of state. We shall put
special emphasis on this fact.

In modern cosmological theories, the cosmological constant remains a focal
point of interest (see \cite{cc1}-\cite{cc4}\ \ for reviews of the problem). A
wide range of observations now compellingly suggest that the universe
possesses a non-zero cosmological constant. Some of the recent discussions on
the cosmological constant \textquotedblleft problem\textquotedblright\ and on
cosmology with a time-varying cosmological constant point out that in the
absence of any interaction with matter or radiation, the cosmological constant
remains a \textquotedblleft constant\textquotedblright. However, in the
presence of interactions with matter or radiation, a solution of Einstein
equations and the assumed equation of covariant conservation of stress-energy
with a time-varying $\Lambda$ can be found. This entails that energy has to be
conserved by a decrease in the energy density of the vacuum component followed
by a corresponding increase in the energy density of matter or radiation
\ Recent observations strongly favour a significant and a positive value of
$\Lambda$ with magnitude $\Lambda(G\hbar/c^{3})\approx10^{-123}$. These
observations suggest an accelerating expansion of the universe, $q<0$.

Therefore the paper is organized as follows. In section II we outline the
metrics as well as the main geometrical ingredients of the models, as for
example, all the curvature invariants. In section III we start by reviewing
some basic notions on self-similarity (SS). We calculate the HVF for the
metric and obtain the restrictions for the scale factors. Section IV is
devoted to revising the \textquotedblleft classical\textquotedblright%
\ solution for a perfect fluid ($\Lambda=0$). Nevertheless we find a new
solution in each of the studied cases i.e. Bianchi types $\mathrm{III}$ and
$\mathrm{VI}$. In section V we shall study the vacuum solution, as in the
above section we obtain solutions for Bianchi types $\mathrm{III}$ and
$\mathrm{VI}$\textrm{. }Once we have finished with the classical solutions we
go next to the study of SS solutions in different contexts. For this purpose,
in section VI we study a perfect fluid model with time-varying constants, $G$
and $\Lambda.$ As we have mentioned above we shall pay special attention to
calculating the numerical value of the equation of state. In section VII we
study a massive cosmic string model while in section VIII we study its
generalization i.e. by considering a model which allows time-varying
constants. In the last section we end by summarizing the main results.

\section{The metrics and the geometric ingredients.}

Throughout the paper $M$ will denote the usual smooth (connected, Hausdorff,
4-dimensional) spacetime manifold with smooth Lorentz metric $g$ of signature
$(-,+,+,+)$ (see for example \cite{MC}). Thus $M$ is paracompact. A comma,
semi-colon and the symbol $\mathcal{L}$ denote the usual partial, covariant
and Lie derivative, respectively, the covariant derivative being with respect
to the Levi-Civita connection on $M$ derived from $g$. The associated Ricci
and stress-energy tensors will be denoted in component form by $R_{ij}(\equiv
R^{c}{}_{jcd})$ and $T_{ij}$ respectively. We shall use a system of units
where $c=1.$

A Bianchi \textrm{VI} space-time is a spatially homogeneous space-time which
admits a group of isometries $G_{3}$, acting on spacelike hypersurfaces,
generated by the spacelike KVs,%
\begin{equation}
\xi_{1}=\partial_{x},\quad\xi_{2}=\partial_{y},\quad\xi_{3}=mx\partial
_{x}-ny\partial_{y}+\partial_{z}, \label{KVF6}%
\end{equation}
and therefore $C_{13}^{1}=m,C_{23}^{2}=-n.$ In synchronous co-ordinates the
metric is:%
\begin{equation}
ds^{2}=-dt^{2}+a^{2}(t)e^{-2mz}dx^{2}+b^{2}(t)e^{2nz}dy^{2}+d^{2}(t)dz^{2},
\label{mB60}%
\end{equation}
where the metric functions $a(t),b(t),d(t),$ are functions of the time
co-ordinate only.

As it is observed, the metric (\ref{mB60}) collapses to the following cases:

\begin{enumerate}
\item if $m=-n,$ then the metric collapses to Bianchi type $\mathrm{V}$,%
\begin{equation}
ds^{2}=-dt^{2}+a^{2}(t)e^{-2mz}dx^{2}+b^{2}(t)e^{-2mz}dy^{2}+d^{2}(t)dz^{2},
\end{equation}
with
\begin{equation}
\xi_{1}=\partial_{x},\quad\xi_{2}=\partial_{y},\quad\xi_{3}=mx\partial
_{x}+my\partial_{y}+\partial_{z}, \label{KVF5}%
\end{equation}
and therefore $C_{13}^{1}=m=C_{23}^{2}.$

\item if $n=0,$ then the metric collapses to Bianchi type $\mathrm{III}$,%
\begin{equation}
ds^{2}=-dt^{2}+a^{2}(t)e^{-2mz}dx^{2}+b^{2}(t)dy^{2}+d^{2}(t)dz^{2},
\end{equation}
with the following KVF (Killings vector fields)%
\begin{equation}
\xi_{1}=\partial_{x},\quad\xi_{2}=\partial_{y},\quad\xi_{3}=mx\partial
_{x}+\partial_{z}, \label{KVF3}%
\end{equation}
and therefore $C_{13}^{1}=m.$

\item If $m=n=0,$ then the model collapses to Bianchi $\mathrm{I}$ model,%
\begin{equation}
ds^{2}=-dt^{2}+a^{2}(t)dx^{2}+b^{2}(t)dy^{2}+d^{2}(t)dz^{2},
\end{equation}
with the following KVF (Killings vector fields).%
\begin{equation}
\xi_{1}=\partial_{x},\quad\xi_{2}=\partial_{y},\quad\xi_{3}=\partial_{z},
\label{KVF1}%
\end{equation}
and therefore $C_{ij}^{k}=0.$
\end{enumerate}

Once we have defined the metric and we know which are its killing vectors,
then we calculate the four velocity. It must verify the equation,
$\mathcal{L}_{\xi_{i}}u_{i}=0,$ so we may define the four velocity as follows:%
\begin{equation}
u^{i}=\left(  1,0,0,0\right)  ,
\end{equation}
in such a way that it is verified, $g(u^{i},u^{i})=-1.$

From the definition of the $4-$velocity we find that:
\begin{align}
H  &  =\left(  \frac{a^{\prime}}{a}+\frac{b^{\prime}}{b}+\frac{d^{\prime}}%
{d}\right)  =\sum_{i}H_{i},\nonumber\\
q  &  =\frac{d}{dt}\left(  \frac{1}{H}\right)  -1,\nonumber\\
\sigma^{2}  &  =\frac{1}{3}\left(  \sum_{i}H_{i}^{2}-\sum_{i\neq j}H_{i}%
H_{j}\right)  .
\end{align}

We shall take into account the Einstein's field equations (FE) written in the
following form:%
\begin{equation}
R_{ij}-\frac{1}{2}Rg_{ij}=8\pi GT_{ij}-\Lambda g_{ij}, \label{EFE}%
\end{equation}
where, $T_{ij},$ is the energy-momentum tensor.

We also study the curvature behaviour of the different solutions (see for
example \cite{Caminati, gron1, gron2} and \cite{Barrow}). The studied
curvature quantities are the following ones. Ricc Scalar, $I_{0}=R_{i}^{i},$
Krestchmann scalar, $I_{1}:=R_{ijkl}R^{ijkl}$, the full contraction of the
Ricci tensor, $I_{2}:=R_{ij}R^{ij}.$ The non-zero components of the Weyl
tensor. The Weyl scalar, $I_{3}=C^{abcd}C_{abcd}=I_{1}-2I_{2}+\frac{1}{3}%
I_{0}^{2},$ as well as the electric scalar $I_{4}=E_{ij}E^{ij},$ (see
\cite{Coley}) and the magnetic scalar $I_{5}=H_{ij}H^{ij},$ of the Weyl
tensor. The Weyl parameter (see W.C. Lim et al \cite{Coley}) which is a
dimensionless measure of the Weyl curvature tensor
\begin{equation}
\mathcal{W}^{2}=\frac{W^{2}}{H^{4}}=\frac{1}{6H^{4}}\left(  E_{ij}%
E^{ij}+H_{ij}H^{ij}\right)  =\frac{I_{3}}{24H^{4}}, \label{Wp}%
\end{equation}
$\mathcal{W}$ can be regarded as describing the intrinsic anisotropy in the
gravitational field. And to end, we shall calculate the gravitational entropy.
From a thermodynamic point of view there is every indication that the entropy
of the universe is \textit{increasing. }Increasing gravitational entropy would
naturally be reflected by increasing local anisotropy, and the Weyl tensor
reflects this. One suggestion in this connection was Penrose's formulation of
what is called the \textit{Weyl curvature conjecture }(WCC) \cite{penrose}.
The hypothesis is motivated by the need for a low entropy constraint on the
initial state of the universe when the matter content was in thermal
equilibrium. Penrose has argued that the low entropy constraint follows from
the existence of the second law of thermodynamics, and that the low entropy in
the gravitational field is tied to constraints on the Weyl curvature.
Wainwright and Anderson \cite{wa} express this conjecture in terms of the
ratio of the Weyl and the Ricci curvature invariants,%
\begin{equation}
P^{2}=\frac{I_{3}}{I_{2}}. \label{GE}%
\end{equation}

The physical content of the conjecture is that the initial state of the
universe is homogeneous and isotropic. As pointed out by Rothman and Anninos
\cite{ra,rothman} the entities $P^{2}$ and $I_{3}$ are \textquotedblleft%
\emph{local\textquotedblright} entities in contrast to what we usually think
of entropy. Gr\o n and Hervik (\cite{gron1,gron2}) have introduced a non-local
entity which shows a more promising behaviour concerning the WCC. This entity
is also constructed in terms of the Weyl tensor, and it has therefore a direct
connection with the Weyl curvature tensor but in a \textquotedblleft%
\emph{non-local form\textquotedblright}.

For SS spacetimes, Pelavas and Lake (\cite{lake}) have pointed out the idea
that eq. (\ref{GE}) is not an acceptable candidate for gravitational entropy
along the homothetic trajectories of any self-similar spacetime. Nor indeed is
any \textquotedblleft dimensionless" scalar. It is showed that the Lie
derivative of any "dimensionless" scalar along a homothetic vector field (HVF)
is zero, and concluded that such functions are not acceptable candidates for
the gravitational entropy. Nevertheless \cite{PC}, since self-similar
spacetimes represent asymptotic equilibrium states (since they describe the
asymptotic properties of more general models), and the result $P^{2}=const.,$
is perhaps consistent with this interpretation since the entropy does not
change in these equilibrium models, and perhaps consequently supports the idea
that $P^{2}$ represents a "gravitational entropy". As we shall show
$\mathcal{W}^{2}$ and $P^{2}$ will be constant along homothetic trajectories,
since all the dimensionless quantities remain constant along timelike
homothetic trajectories.

\section{The Self-similar solution.}

In general relativity, the term self-similarity can be used in two ways. One
is for the properties of space-times, the other is for the properties of
matter fields. These are not equivalent in general. The self-similarity in
general relativity was defined for the first time by Cahill and Taub
\cite{CT}, Eardley \cite{22} (see for general reviews \cite{CC}-\cite{Hall}).
Self-similarity is defined by the existence of a homothetic vector ${V}$ in
the spacetime, which satisfies
\begin{equation}
\mathcal{L}_{V}g_{ij}=2\alpha g_{ij}, \label{gss1}%
\end{equation}
where $g_{ij}$ is the metric tensor, $\mathcal{L}_{V}$ denotes Lie
differentiation along ${V}$ and $\alpha$ is a constant. This is a special type
of conformal Killing vectors. This self-similarity is called homothety. If
$\alpha\neq0$, then it can be set to be unity by a constant rescaling of ${V}%
$. If $\alpha=0$, i.e. $\mathcal{L}_{V}g_{ij}=0$, then ${V}$ is a Killing vector.

Homothety is a purely geometric property of spacetime so that the physical
quantity does not necessarily exhibit self-similarity such as $\mathcal{L}%
_{V}Z=kZ$, where $k$ is a constant and $Z$ is, for example, the pressure, the
energy density and so on. From equation (\ref{gss1}) it follows that
$\mathcal{L}_{V}R^{i}\,_{jkl}=0,$ and hence $\mathcal{L}_{V}R_{ij}=0,$and
$\mathcal{L}_{V}G_{ij}=0.$ A vector field ${V}$ that satisfies the above
equations is called a curvature collineation, a Ricci collineation and a
matter collineation, respectively. It is noted that such equations do not
necessarily mean that ${V}$ is a homothetic vector. We consider the Einstein
equations $G_{ij}=8\pi GT_{ij},$ where $T_{ij}$ is the energy-momentum tensor.
If the spacetime is homothetic, the energy-momentum tensor of the matter
fields must satisfy $\mathcal{L}_{V}T_{ij}=0$. For a perfect fluid case, the
energy-momentum tensor takes the form $T_{ij}=(p+\rho)u_{i}u_{j}+pg_{ij}%
,$where $p$ and $\rho$ are the pressure and the energy density, respectively.
Then, equations~(\ref{gss1}) result in
\begin{equation}
\mathcal{L}_{V}u^{i}=-\alpha u^{i},\quad\mathcal{L}_{V}\rho=-2\alpha\rho
,\quad\mathcal{L}_{V}p=-2\alpha p. \label{ssmu}%
\end{equation}

As shown above, for a perfect fluid, the self-similarity of the spacetime and
that of the physical quantity coincide. However, this fact does not
necessarily hold for more general matter fields. Thus the self-similar
variables can be determined from dimensional considerations in the case of
homothety. Therefore, we can conclude homothety as the general relativistic
analogue of complete similarity. From the constraints (\ref{ssmu}), we can
show that if we consider the barotropic equation of state, i.e., $p=f(\rho)$,
then the equation of state must have the form $p=\omega\rho$, where $\omega$
is a constant. In this paper, we would like to try to show how taking into
account this class of hypothesis one is able to find exact solutions to the
field equations within the framework of the time varying constants.

From equation (\ref{gss1})$,$we find the following homothetic vector field for
the B$\mathrm{VI}_{0}$ metric (\ref{mB60}):%
\begin{equation}
V=t\partial_{t}+\left(  1-t\frac{a^{\prime}}{a}\right)  x\partial_{x}+\left(
1-t\frac{b^{\prime}}{b}\right)  y\partial_{y}+\left(  1-t\frac{d^{\prime}}%
{d}\right)  z\partial_{z},
\end{equation}
with the following constrains for the scale factors:
\begin{equation}
a(t)=a_{0}t^{a_{1}},\qquad b(t)=b_{0}t^{a_{2}},\qquad d(t)=d_{0}%
t,\label{restric}%
\end{equation}
where $a_{1},a_{2}\in\mathbb{R}$ and we set $d_{0}=1$, therefore the resulting
homothetic vector field is:%
\begin{equation}
V=t\partial_{t}+\left(  1-a_{1}\right)  x\partial_{x}+\left(  1-a_{2}\right)
y\partial_{y},
\end{equation}
so the metric (\ref{mB60}) collapses to the following one:%
\begin{equation}
ds^{2}=-dt^{2}+a^{2}(t)e^{-2mz}dx^{2}+b^{2}(t)e^{2nz}dy^{2}+t^{2}dz^{2}.
\end{equation}

\section{Perfect fluid solutions.}

The energy-momentum tensor for a perfect fluid model$,$ $T_{ij},is$ defined as
follows:%
\begin{equation}
T_{ij}=(\rho+p)u_{i}u_{j}+pg_{ij}, \label{eq11}%
\end{equation}
with the usual equation of state $p=\omega\rho,$ $\omega\in\mathbb{R}.$ The
resulting FE for the metric (\ref{mB60}) with a perfect fluid model yield
(with $\Lambda=0$):
\begin{align}
\frac{a^{\prime}}{a}\frac{b^{\prime}}{b}+\frac{a^{\prime}}{a}\frac{d^{\prime}%
}{d}+\frac{d^{\prime}}{d}\frac{b^{\prime}}{b}+\frac{K}{d^{2}}  &  =8\pi
G\rho,\label{fe1}\\
m\left(  \frac{a^{\prime}}{a}-\frac{d^{\prime}}{d}\right)  +n\left(
\frac{d^{\prime}}{d}-\frac{b^{\prime}}{b}\right)   &  =0,\label{fe2}\\
\frac{b^{\prime\prime}}{b}+\frac{d^{\prime\prime}}{d}+\frac{d^{\prime}}%
{d}\frac{b^{\prime}}{b}-\frac{n^{2}}{d^{2}}  &  =-8\pi G\omega\rho
,\label{fe3}\\
\frac{d^{\prime\prime}}{d}+\frac{a^{\prime\prime}}{a}+\frac{a^{\prime}}%
{a}\frac{d^{\prime}}{d}-\frac{m^{2}}{d^{2}}  &  =-8\pi G\omega\rho
,\label{fe4}\\
\frac{b^{\prime\prime}}{b}+\frac{a^{\prime\prime}}{a}+\frac{a^{\prime}}%
{a}\frac{b^{\prime}}{b}+\frac{mn}{d^{2}}  &  =-8\pi G\omega\rho,\label{fe5}\\
\rho^{\prime}+\rho\left(  1+\omega\right)  \left(  \frac{a^{\prime}}{a}%
+\frac{b^{\prime}}{b}+\frac{d^{\prime}}{d}\right)   &  =0, \label{fe6}%
\end{align}
where $K=\left(  mn-m^{2}-n^{2}\right)  .$

Therefore, taking into account the restrictions for the scale factors given by
eq. (\ref{restric}), the solution for a perfect fluid model, eqs.
(\ref{fe1}-\ref{fe6}), is the following one.

Form (\ref{fe6}) we get:%
\[
\rho=\rho_{0}t^{-\left(  1+\omega\right)  \left(  a_{1}+a_{2}+1\right)  }%
=\rho_{0}t^{-\gamma},
\]
with $\left(  1+\omega\right)  \left(  a_{1}+a_{2}+1\right)  =\left(
1+\omega\right)  \alpha=\gamma.$ Now, taking into account eq. (\ref{fe1}) we
get $a_{1}a_{2}+a_{1}+a_{2}+K=8\pi G\rho_{0},$ and $\left(  1+\omega\right)
\left(  a_{1}+a_{2}+1\right)  =2,$ and hence (\ref{fe1}-\ref{fe6}) collapse to
the following algebraic system of equations:
\begin{align}
m\left(  a_{1}-1\right)  +n\left(  1-a_{2}\right)   &  =0,\label{sys1}\\
a_{2}\left(  a_{2}-1\right)  +a_{2}-n^{2}  &  =-A\omega,\label{sys2}\\
a_{1}\left(  a_{1}-1\right)  +a_{1}-m^{2}  &  =-A\omega,\label{sys3}\\
a_{2}\left(  a_{2}-1\right)  +a_{1}\left(  a_{1}-1\right)  +a_{1}a_{2}+mn  &
=-A\omega,\label{sys4}\\
\left(  1+\omega\right)  \left(  a_{1}+a_{2}+1\right)   &  =2, \label{sys5}%
\end{align}
with%
\[
\rho_{0}=\frac{A}{8\pi G},\qquad A=a_{1}a_{2}+a_{1}+a_{2}+K,
\]
obtaining the following set of solutions.

\subsection{Bianchi type III.}

Our first solution is the following one:%
\begin{equation}
a_{1}=1,\,\,a_{2}=-\frac{2\omega}{\omega+1},\,\,m=\frac{\sqrt{-3\omega
^{2}+2\omega+1}}{\omega+1},\,\,n=0,
\end{equation}
note that the solution only has sense if $\omega\in\left(  -\frac{1}%
{3},1\right)  ,$ and therefore we obtain the following values: $a_{1}=1,$
$a_{2}\in\left(  0,1\right)  ,$ iff $\omega<0,$ $m\in\left(  0,1\right)  ,$and
$\left(  1+\omega\right)  \left(  2+a_{2}\right)  =\gamma=2,$ so $\rho
=\rho_{0}t^{-2},$ and therefore the metric collapses to the following one%
\begin{equation}
ds^{2}=-dt^{2}+t^{2}e^{-2mz}dx^{2}+t^{2a_{2}}dy^{2}+t^{2}dz^{2},
\label{sol1ss}%
\end{equation}
which stand for a Bianchi $\mathrm{III}$ metric. Note that (\ref{sol1ss})
admits the KVF given by eq. (\ref{KVF3}). This solution looks really
restrictive since it only has physical meaning if $\omega<0.$

For this solution we have obtained the following behaviour for the main
curvature quantities. We start by calculating the curvature invariants,
$\left(  I_{i}\right)  _{i=0}^{2},$ they yield: $I_{0}=2\left(  a_{2}%
^{2}+a_{2}-m^{2}+1\right)  t^{-2},$ $I_{1}=4\left(  a_{2}^{4}-2a_{2}%
^{3}+3a_{2}^{2}+m^{4}-2m^{2}+1\right)  t^{-4},$ and $I_{2}=2\left(  a_{2}%
^{4}+2a_{2}^{2}+2a_{2}+m^{4}-2m^{2}+1-2m^{2}a_{2}\right)  t^{-4}.$ As it is
observed we have obtained a singular solution. The non-zero components of the
Weyl tensor are: $C_{txtx}=-(1/6)\mathcal{K}e^{-2mz},$ $C_{tyty}%
=(1/3)\mathcal{K}t^{2\left(  a_{2}-1\right)  },$ $C_{tztz}=-(1/6)\mathcal{K},$
$C_{xyxy}=(1/6)\mathcal{K}e^{-2mz}t^{2a_{2}},$ $C_{xzxz}=-(1/3)\mathcal{K}%
e^{-2mz}t^{2},$ and $C_{yzyz}=(1/6)\mathcal{K}t^{2a_{2}},$ where
$\mathcal{K=}\left(  m-1+a_{2}\right)  \left(  m+1-a_{2}\right)  .$ Therefore
the Weyl invariant yields, $I_{3}=\left(  4/3\right)  \mathcal{K}^{2}t^{-4}.$
The electric invariant yields: $I_{4}=\left(  1/6\right)  \mathcal{K}%
^{2}t^{-4},$ while the magnetic invariant, $I_{5},$ vanishes. The Weyl
parameter yields, $\mathcal{W}^{2}=\mathcal{K}^{2}/36(2+a_{2})^{4}%
=\mathrm{const}$\textrm{.,} while the gravitational entropy yields,
$P^{2}=\mathrm{const}$\textrm{.,} as it is expected\textrm{.}

\subsection{Bianchi type $\mathrm{VI}_{0}$, solution I.}

The classical solution given by Collins \cite{Collins}, Hsu et al (see
\cite{HW}), i.e.
\begin{equation}
a_{1}=a_{2}=\frac{1-\omega}{2\left(  \omega+1\right)  },\,\,m=n=\frac
{\sqrt{-3\omega^{2}+2\omega+1}}{2\left(  \omega+1\right)  },
\end{equation}
with $\omega\in\left(  -\frac{1}{3},1\right)  ,$ and therefore the metric
collapses to the following one%
\begin{equation}
ds^{2}=-dt^{2}+t^{2a_{1}}e^{-2mz}dx^{2}+t^{2a_{1}}e^{2mz}dy^{2}+t^{2}dz^{2},
\label{B6c}%
\end{equation}
finding in this way that, $a_{2}\in\left(  0,1\right)  ,$ $m\in\left(
0,\frac{1}{2}\right)  $ and $\left(  1+\omega\right)  \left(  1+2a_{2}\right)
=\gamma=2,$ so $\rho=\rho_{0}t^{-2}.$ In the way we find that%
\[
H=\frac{1+2a_{2}}{t},\quad q=\frac{-2a_{2}}{1+2a_{2}}<0,\quad\sigma^{2}%
=\frac{\left(  a_{2}-1\right)  ^{2}}{3t^{2}}.
\]

For this solution we have obtained the following behaviour for the main
curvature quantities. We start by calculating the curvature invariants,
$\left(  I_{i}\right)  _{i=0}^{2},$ they yield: $I_{0}=2\left(  3a_{1}%
^{2}-m^{2}\right)  t^{-2},$ $I_{1}=4\left(  3a_{1}^{4}-4a_{1}^{3}+4a_{1}%
^{2}-2a_{1}^{2}m^{2}+3m^{4}+4m^{2}a_{1}-4m^{2}\right)  t^{-4},$ and
$I_{2}=2\left(  3a_{1}^{4}-2a_{1}^{3}+2a_{1}^{2}+m^{4}-2m^{2}a_{1}\right)
t^{-4}.$ As it is observed we have obtained a singular solution, as it is
expect for a SS solution. The non-zero components of the Weyl tensor are:
$C_{txtx}=(1/3)m^{2}t^{2\left(  a_{1}-1\right)  }e^{-2mz},$ $C_{txxz}=m\left(
1-a_{1}\right)  t^{2a_{1}-1}e^{-2mz},$ $C_{tyty}=(1/3)m^{2}t^{2\left(
a_{1}-1\right)  }e^{2mz},$ $C_{tyyz}=m\left(  a_{1}-1\right)  t^{2a_{1}%
-1}e^{2mz},$ $C_{tztz}=-(2/3)m^{2},$ $C_{xyxy}=(2/3)m^{2}t^{2\left(
2a_{1}-1\right)  },$ $C_{xzxz}=-(1/3)t^{2a_{1}}e^{-2mz},$ and $C_{yzyz}%
=(1/3)t^{2a_{1}}e^{2mz}.$ Therefore the Weyl invariant yields, $I_{3}=\left(
16/3\right)  \left(  -3a_{1}^{2}+6a_{1}-3+m^{2}\right)  t^{-4}.$ The electric
invariant yields: $I_{4}=\left(  2/3\right)  m^{4}t^{-4},$ while the magnetic
invariant, $I_{5},$ behaves as: $I_{5}=2m^{2}\left(  a_{1}-1\right)
^{2}t^{-4}$. The Weyl parameter yields, $\mathcal{W}^{2}=\left(  m^{2}\left(
3a_{1}^{2}-6a_{1}+3+m^{2}\right)  \right)  /9(1+2a_{1})^{4}=\mathrm{const.}%
$\textrm{,} while the gravitational entropy yields, $P^{2}=\mathrm{const.}%
$\textrm{,} as it is expected\textrm{.}

\subsection{Bianchi type $\mathrm{VI}_{0}$, solution II.}

We find a new solution, where
\begin{align}
a_{1}  &  =\frac{1-a_{2}\left(  \omega+1\right)  -\omega}{\omega+1}%
,\,\,a_{2}=a_{2},\nonumber\\
n  &  =\frac{b\left(  a_{2}\left(  \omega+1\right)  +2\omega\right)  }{\left(
\omega+1\right)  },\,\,m=b\left(  1-a_{2}\right)  ,\\
b  &  =\sqrt{\frac{1-\omega}{3\omega+1}},\nonumber
\end{align}
and at a first look we may say that $\omega\in(-\frac{1}{3},1],$ and therefore
$b\in\left[  0,\infty\right)  .$

The metric collapses to the following one%
\begin{equation}
ds^{2}=-dt^{2}+t^{2a_{1}}e^{-2mz}dx^{2}+t^{2a_{2}}e^{2nz}dy^{2}+t^{2}dz^{2},
\end{equation}
finding in this way that, $a_{1}=a_{1}\left(  a_{2},\omega\right)  ,$ so
depending on the different values of $\omega$ we get:%
\[%
\begin{array}
[c]{|c|c|c|c|c|}\hline
\omega & a_{1} & a_{2} & n & m\\\hline\hline
1 & -a_{2} & a_{2} & 0 & 0\\\hline
1/3 & \frac{1}{2}-a_{2} & \left(  0,1/2\right)  & \frac{\sqrt{3}}{6}\left(
2a_{2}+1\right)  & \frac{\sqrt{3}}{3}\left(  1-a_{2}\right) \\\hline
0 & 1-a_{2} & \left(  0,1\right)  & a_{2} & 1-a_{2}\\\hline
\rightarrow\left(  -1/3\right)  ^{+} & 2-a_{2} & \left(  0,2\right)  & n &
m\\\hline
\end{array}
\]

Note that the case $\omega=1$ looks nonphysical since the scale factor
$a_{1}<0,$ for this reason we shall not take into account this particular
case. Therefore the solution is only valid if $\omega\in\left(  -\frac{1}%
{3},1\right)  $ and hence we get $\left(  1+\omega\right)  \left(
1+a_{1}+a_{2}\right)  =\gamma=2,$ so $\rho=\rho_{0}t^{-2}.$ In the way we find
that%
\[
H=\frac{1+a_{1}+a_{2}}{t},\quad q=\frac{-\left(  a_{1}+a_{2}\right)  }%
{1+a_{1}+a_{2}}<0,
\]
and
\[
\sigma^{2}=\frac{2}{3t^{2}}\left(  a_{1}^{2}+a_{2}^{2}+1-a_{1}a_{2}%
-a_{1}-a_{2}\right)  .
\]

For simplicity we calculate the curvature behaviour for the case $\omega=0.$
We start by calculating the curvature invariants, $\left(  I_{i}\right)
_{i=0}^{2},$ they yield: $I_{0}=4a_{2}\left(  a_{2}-1\right)  t^{-2},$
$I_{1}=16a_{2}^{2}\left(  a_{2}-1\right)  ^{2}t^{-4}=I_{2}.$ As it is observed
we have obtained a singular solution, as it is expect for a SS solution. The
non-zero components of the Weyl tensor are: $C_{txtx}\sim t^{-2a_{2}},$
$C_{txxz}\sim t^{1-2a_{2}},$ $C_{tyty}\sim t^{2\left(  a_{2}-1\right)  },$
$C_{tyyz}\sim t^{2a_{2}-1},$ $C_{tztz}=\mathrm{const.},$ $C_{xyxy}%
=\mathrm{const.},$ $C_{xzxz}\sim t^{2\left(  1-a_{2}\right)  },$ and
$C_{yzyz}\sim t^{2a_{2}}.$ Therefore the Weyl invariant yields, $I_{3}%
=32a_{2}^{2}\left(  a_{2}-1\right)  ^{2}/3t^{4}.$ The electric invariant
yields: $I_{4}=2a_{2}^{2}\left(  a_{2}-1\right)  ^{2}/3t^{4},$ while the
magnetic invariant, $I_{5},$ behaves as: $I_{5}=2a_{2}^{2}\left(
a_{2}-1\right)  ^{2}t^{-4}$. The Weyl parameter yields, $\mathcal{W}^{2}%
=a_{2}^{2}\left(  a_{2}-1\right)  ^{2}/36=\mathrm{const.}$\textrm{,} while the
gravitational entropy yields, $P^{2}=2/3$\textrm{.,} as it is
expected\textrm{.}

\section{SS vacuum solution.}

In this case, the system to solve is the following one:%
\begin{align}
a_{1}a_{2}+a_{1}+a_{2}+\left(  mn-m^{2}-n^{2}\right)   &  =0,\\
m\left(  a_{1}-1\right)  +n\left(  1-a_{2}\right)   &  =0,\\
a_{2}\left(  a_{2}-1\right)  +a_{2}-n^{2}  &  =0,\\
a_{1}\left(  a_{1}-1\right)  +a_{1}-m^{2}  &  =0,\\
a_{2}\left(  a_{2}-1\right)  +a_{1}\left(  a_{1}-1\right)  +a_{1}a_{2}+mn  &
=0,
\end{align}
obtaining the following solutions (as well as the trivial one).

\begin{enumerate}
\item We find our first solution as:%
\begin{align}
a_{2}  &  =a_{2},\qquad n=-a_{2},\\
a_{1}  &  =\frac{1}{2}\left(  1+\sqrt{-2a_{2}^{2}+4a_{2}+1}\right)  =m,
\end{align}
and where the solution only has math sense if
\[
a_{2}\in\left(  1-\frac{1}{2}\sqrt{6},1+\frac{1}{2}\sqrt{6}\right)
\]
the negative solution may have sense in the study of singularities. We only
consider the positive range i.e. $a_{2}\in\left(  0,2.2247\right)  ,$ and
therefore we get, $a_{1}\in\left(  0,1,35\right)  .$

Equivalently we find%
\begin{align*}
a_{2}  &  =a_{2},\qquad n=a_{2},\\
a_{1}  &  =\frac{1}{2}\left(  1+\sqrt{-2a_{2}^{2}+4a_{2}+1}\right)  =-m,
\end{align*}
so the metric collapses to the following one%
\begin{equation}
ds^{2}=-dt^{2}+t^{2a_{1}}e^{2a_{1}z}dx^{2}+t^{2a_{2}}e^{2a_{2}z}dy^{2}%
+t^{2}dz^{2}, \label{B5}%
\end{equation}
it describes a metric belonging to Bianchi type \textrm{VI} (i.e. it admits
the KVF given by
\[
\xi_{1}=\partial_{x},\quad\xi_{2}=\partial_{y},\quad\xi_{3}=-a_{1}%
x\partial_{x}-a_{2}\partial_{y}+\partial_{z},
\]
and hence $C_{13}^{1}=-a_{1},C_{23}^{2}=-a_{2}$, and to the best of our
knowledge is new.

For this solution we have found a really curious curvature behaviour. The
curvature invariants $\left(  I_{i}\right)  _{i=0}^{2},$ vanish. The non-zero
components of the Weyl tensor are: $C_{txtx}=-\left(  1/2\right)
\mathcal{K}e^{2a_{1}z}t^{2(a_{1}-1)},$ $C_{txxz}=\left(  1/2\right)
\mathcal{K}e^{2a_{1}z}t^{2a_{1}-1},$ $C_{tyty}=\left(  1/2\right)
\mathcal{K}e^{2a_{2}z}t^{2(a_{2}-1)},$ $C_{tyyz}=-\left(  1/2\right)
\mathcal{K}e^{2a_{2}z}t^{2a_{2}-1},$ $C_{xzxz}=-\left(  1/2\right)
\mathcal{K}e^{2a_{1}z}t^{2a_{1}},$ and $C_{yzyz}=\left(  1/2\right)
\mathcal{K}e^{2a_{2}z}t^{2a_{2}},$ where $\mathcal{K}=(a_{1}-1+a_{2}%
)(a_{1}-a_{2}).$ Therefore the Weyl invariant yields, $I_{3}=0.$ The electric
invariant yields: $I_{4}=\left(  1/2\right)  \mathcal{K}^{2}t^{-4},$ while the
magnetic invariant, $I_{5},$ behaves as: $I_{5}=\left(  1/2\right)
\mathcal{K}^{2}t^{-4}$. The Weyl parameter yields, $\mathcal{W}^{2}%
=\mathcal{K}^{2}/\left(  6\left(  a_{1}+a_{2}+1\right)  ^{4}\right)
=\mathrm{const.}$\textrm{,} while the gravitational entropy yields,
$P^{2}=\mathrm{\infty}$\textrm{,} as it is expected, since $I_{2}=0$\textrm{.}

\item And the set of solutions:%
\begin{align*}
a_{2}  &  =n=0,\qquad a_{1}=m=1,\\
a_{2}  &  =n=1,\qquad a_{1}=m=0,\qquad and\\
a_{2}  &  =n=0,\qquad a_{1}=1,\qquad m=-1,
\end{align*}
so the metric collapses to the following one (see Hsu et al. \cite{HW} section
2.6.3)%
\begin{equation}
ds^{2}=-dt^{2}+t^{2}e^{2z}dx^{2}+dy^{2}+t^{2}dz^{2}. \label{B3d}%
\end{equation}
This solution describes a Bianchi $\mathrm{III}$ metric, admitting three KVF
given by eq. (\ref{KVF3}) with $m=-1$. For this solution we have found a
really pathological behaviour. All the curvature quantities vanish. So this
solution lacks of any physical interest.
\end{enumerate}

\section{Perfect fluid model with variable constants.}

The resulting FE for the metric (\ref{mB60}) with a perfect fluid matter model
(\ref{eq11}) and with $G$ and $\Lambda$ time-varying are:%
\begin{align}
\frac{a^{\prime}}{a}\frac{b^{\prime}}{b}+\frac{a^{\prime}}{a}\frac{d^{\prime}%
}{d}+\frac{d^{\prime}}{d}\frac{b^{\prime}}{b}+\frac{K}{d^{2}}  &  =8\pi
G\rho+\Lambda,\label{I1}\\
m\left(  \frac{a^{\prime}}{a}-\frac{d^{\prime}}{d}\right)  +n\left(
\frac{d^{\prime}}{d}-\frac{b^{\prime}}{b}\right)   &  =0,\label{I2}\\
\frac{b^{\prime\prime}}{b}+\frac{d^{\prime\prime}}{d}+\frac{d^{\prime}}%
{d}\frac{b^{\prime}}{b}-\frac{n^{2}}{d^{2}}  &  =-8\pi G\omega\rho
+\Lambda,\label{I3}\\
\frac{d^{\prime\prime}}{d}+\frac{a^{\prime\prime}}{a}+\frac{a^{\prime}}%
{a}\frac{d^{\prime}}{d}-\frac{m^{2}}{d^{2}}  &  =-8\pi G\omega\rho
+\Lambda,\label{I4}\\
\frac{b^{\prime\prime}}{b}+\frac{a^{\prime\prime}}{a}+\frac{a^{\prime}}%
{a}\frac{b^{\prime}}{b}+\frac{mn}{d^{2}}  &  =-8\pi G\omega\rho+\Lambda
,\label{I5}\\
\rho^{\prime}+\rho\left(  1+\omega\right)  H  &  =0,\label{I6}\\
\Lambda^{\prime}  &  =-8\pi G^{\prime}\rho. \label{I7}%
\end{align}

Now, we shall take into account the obtained SS restrictions for the scale
factors given by eq. (\ref{restric}).

From eq. (\ref{I6}) we get%
\begin{equation}
\rho=\rho_{0}t^{-\gamma}, \label{t_den}%
\end{equation}
where $\gamma=\left(  \omega+1\right)  \alpha$ and $\alpha=\left(  a_{1}%
+a_{2}+1\right)  $.

From eq. (\ref{I1}) we obtain:%
\begin{equation}
\Lambda=\left[  At^{-2}-8\pi G\rho_{0}t^{-\left(  \omega+1\right)  \alpha
}\right]  , \label{peta1}%
\end{equation}
where $A=a_{1}a_{2}+a_{1}+a_{2}+K,$ and $K=\left(  mn-m^{2}-n^{2}\right)  .$

Now, taking into account eq. (\ref{I7}) and eq. (\ref{peta1}), algebra brings
us to obtain%
\begin{equation}
G=G_{0}t^{\gamma-2},\qquad G_{0}=\frac{A}{4\pi\rho_{0}\left(  \omega+1\right)
\alpha}. \label{G}%
\end{equation}

While the cosmological \textquotedblleft constant\textquotedblright\ behaves
as:
\begin{equation}
\Lambda=\frac{A}{c^{2}}\left(  1-\frac{2}{\gamma}\right)  t^{-2}=\Lambda
_{0}t^{-2}.
\end{equation}

With all these result we find that the system to solve is the following one:%
\begin{align}
m\left(  a_{1}-1\right)  +n\left(  1-a_{2}\right)   &  =0,\\
a_{2}(a_{2}-1)+a_{2}-n^{2}  &  =\mathcal{A},\\
a_{1}\left(  a_{1}-1\right)  +a_{1}-m^{2}  &  =\mathcal{A}\\
a_{2}(a_{2}-1)+a_{1}a_{2}+a_{1}(a_{1}-1)+mn  &  =\mathcal{A},
\end{align}
where $\mathcal{A}=A\left(  \frac{\alpha-2}{\alpha}\right)  $ whose solutions
are the following ones.

\subsection{Bianchi type $\mathrm{VI}_{0}$.}

As it is observed we obtain a solution that is valid for all equation of state
$\omega,$ with%
\[
a_{2}=a_{2},\qquad a_{1}=\frac{a_{2}^{2}+1}{a_{2}+1},
\]%
\[
m=\sqrt{\frac{a_{2}^{2}+1-2a_{2}^{3}}{2a_{2}+1}},\quad n=\frac{ma_{2}}%
{a_{2}+1}.
\]

So this solution only has physical meaning if $a_{2}\in\left(  0,1\right)  ,$
and therefore we have : $a_{1}\in\left(  0.82,1\right)  ,$ $m\in\left(
0,1\right)  ,$ $n\in\left(  0,1\right)  ,$ so the metric takes the following
form:
\begin{equation}
ds^{2}=-dt^{2}+t^{2a_{1}}e^{-2mz}dx^{2}+t^{2a_{2}}e^{2nz}dy^{2}+t^{2}dz^{2}.
\end{equation}

Therefore we have the following behaviour for the main quantities:.
$\gamma=\left(  \omega+1\right)  \alpha\in\left(  0.6,6\right)  ,$
$\alpha=\left(  a_{1}+a_{2}+1\right)  \in\left(  1.82,3\right)  ,$ and
$A=a_{1}a_{2}+a_{1}+a_{2}+K>0,$ therefore%
\[
\rho=\rho_{0}t^{-\gamma},\qquad\gamma\in\left(  0.6,6\right)  ,
\]
is a positive decreasing time function.

$G$ behaves as:%
\[
G=G_{0}t^{\gamma-2},
\]
so it will be a growing function if $\gamma>2$, constant if $\gamma=2$ and a
decreasing time function if $\gamma<2.$

The cosmological constant behaves as%
\[
\Lambda=A\left(  1-\frac{2}{\gamma}\right)  t^{-2},
\]
hence its sign will depend on the value of $\gamma.$ If $\gamma<2$ then we get
a negative $\Lambda$, it will behave as a true constant if $\gamma=2,$ and we
find that it is a positive time decreasing function if $\gamma>2.$ Note that
if $\gamma>2,$ then $G$ is growing and $\Lambda_{0}>0.$

The rest of the quantities behave as%
\begin{align*}
H  &  =\left(  a_{1}+a_{2}+1\right)  \frac{1}{t}=\frac{\alpha}{t},\\
q  &  =\frac{d}{dt}\left(  \frac{1}{H}\right)  -1=\frac{1-\alpha}{\alpha
}<0,\quad\forall\alpha,\\
\sigma^{2}  &  =\frac{1}{3}\left(  a_{1}^{2}+a_{2}^{2}+1-a_{1}a_{2}%
-a_{1}-a_{2}\right)  .
\end{align*}
For example we may see in the following table the behaviour of the main
quantities for a particular value of $a_{2}.$%
\[%
\begin{array}
[c]{|c|c|c|c|c|c|}\hline
a_{2} & a_{1} & \omega & \gamma & G & \Lambda\\\hline\hline
1/2 & 0.833 & 1 & >2 & \nearrow & >0\\\hline
1/2 & 0.833 & 1/3 & >2 & \nearrow & >0\\\hline
1/2 & 0.833 & 0 & >2 & \nearrow & >0\\\hline
1/2 & 0.833 & -1/3 & <2 & \searrow & <0\\\hline
\end{array}
\]
choosing other parameters we find another behaviour.

For this solution we have obtained the following behaviour for the main
curvature quantities. We start by calculating the curvature invariants,
$\left(  I_{i}\right)  _{i=0}^{2},$ they yield: $I_{0}=2K_{0}t^{-2},$
$I_{1}=4K_{1}t^{-4},$ and $I_{2}=2K_{2}t^{-4},$ where $K_{i}=K\left(
a_{1},a_{2},m,n\right)  $ are numerical constants. As it is observed we have
obtained a singular solution, as it is expect for a SS solution. The non-zero
components of the Weyl tensor are: $C_{txtx}=\mathcal{K}_{1}t^{2\left(
a_{1}-1\right)  }e^{-2mz},$ $C_{txxz}=\mathcal{K}_{2}t^{2a_{1}-1}e^{-2mz},$
$C_{tyty}=\mathcal{K}_{3}t^{2\left(  a_{1}-1\right)  }e^{2nz},$ $C_{tyyz}%
=\mathcal{K}_{4}t^{2a_{1}-1}e^{2nz},$ $C_{tztz}=\mathcal{K}_{5},$
$C_{xyxy}=\mathcal{K}_{6}e^{2z\left(  n-m\right)  }t^{2\left(  a_{1}%
+a_{2}+1\right)  },$ $C_{xzxz}=\mathcal{K}_{7}t^{2a_{1}}e^{-2mz},$ and
$C_{yzyz}=\mathcal{K}_{8}t^{2a_{2}}e^{2nz},$ where $\mathcal{K}_{i}%
=\mathcal{K}\left(  a_{1},a_{2},m,n\right)  $ are numerical constants.
Therefore the Weyl invariant yields, $I_{3}=\left(  4/3\right)  K_{3}t^{-4}.$
The electric invariant yields: $I_{4}=\left(  1/6\right)  K_{4}t^{-4},$ while
the magnetic invariant, $I_{5},$ behaves as, $I_{5}=\left(  1/6\right)
K_{5}t^{-4}$. The Weyl parameter yields, $\mathcal{W}^{2}=\mathrm{const.}%
$\textrm{,} (note that $\mathcal{W}^{2}\rightarrow0$ is really small, so the
model isotropize) while the gravitational entropy yields, $P^{2}%
=\mathrm{const.}$\textrm{,} as it is expected\textrm{.}

\subsection{Bianchi $\mathrm{III}$ solution.}

This solution is a Bianchi $\mathrm{III}$ metric with:%
\[
a_{1}=a_{1},\quad a_{2}=1,\quad m=0,\quad n=\sqrt{1-a_{1}^{2}},
\]
so $a_{1}\in\left(  0,1\right)  .$ In this way the metric collapses to the
following one:%
\begin{equation}
ds^{2}=-dt^{2}+t^{2a_{1}}dx^{2}+t^{2}\left(  e^{2nz}dy^{2}+dz^{2}\right)  ,
\end{equation}
note that if $a_{1}\rightarrow1,$ then we get
\[
ds^{2}=-dt^{2}+t^{2}\left(  dx^{2}+dy^{2}+dz^{2}\right)  ,
\]
this is fact the self-similar solution obtained from a Bianchi $\mathrm{V}$
metric, which is only valid for $\omega=-1/3,$ (in fact, the SS Bianchi
$\mathrm{V},$ is the FRW solution with $k=-1,$ and $\omega=-1/3.)$

With regard to the main quantities we find that.%
\begin{align*}
H  &  =\left(  a_{1}+2\right)  \frac{1}{t}=\frac{\alpha}{t},\\
q  &  =\frac{d}{dt}\left(  \frac{1}{H}\right)  -1=\frac{1-\alpha}{\alpha},\\
\sigma^{2}  &  =\frac{1}{3}\left(  a_{1}-1\right)  ^{2}.
\end{align*}
Hence we have the following behaviour for the main quantities: $\gamma=\left(
\omega+1\right)  \alpha\in\left(  1,6\right)  ,$ $\alpha=\left(
a_{1}+2\right)  \in\left(  2,3\right)  ,$ and $\ A=a_{1}a_{2}+a_{1}%
+a_{2}+K>0,$ therefore%
\[
\rho=\rho_{0}t^{-\gamma},\qquad\gamma\in\left(  1,6\right)  ,
\]
is a positive decreasing time function. $\gamma<2\Longleftrightarrow\omega<0.$

$G$ behaves as:%
\[
G=G_{0}t^{\gamma-2},
\]
so it will be a growing function if $\gamma>2$, constant if $\gamma=2$ and a
decreasing time function if $\gamma<2.$

The cosmological constant behaves as%
\[
\Lambda=A\left(  1-\frac{2}{\gamma}\right)  t^{-2},
\]
hence its sign will depend on the value of $\gamma.$ If $\gamma<2$ then we get
a negative $\Lambda$, it will behave as a true constant if $\gamma=2$ and we
find a positive time decreasing function if $\gamma>2.$

For this solution we have obtained the following behaviour for the main
curvature quantities. We start by calculating the curvature invariants,
$\left(  I_{i}\right)  _{i=0}^{2},$ they yield: $I_{0}=2\left(  a_{1}%
^{2}+a_{1}+1-n^{2}\right)  t^{-2},$ $I_{1}=4\left(  3a_{1}^{4}-2a_{1}%
^{3}+2a_{1}^{2}+n^{4}-2n^{2}+1\right)  t^{-4},$ and $I_{2}=2\left(  a_{1}%
^{4}+2a_{1}^{2}+2a_{1}+n^{4}-2n^{2}a_{1}-2n^{2}+1\right)  t^{-4}.$ As it is
observed we have obtained a singular solution, as it is expect for a SS
solution. The non-zero components of the Weyl tensor are the following ones:
$C_{txtx}=(1/3)\mathcal{K}t^{2\left(  a_{1}-1\right)  },$ $C_{tyty}%
=-(1/6)\mathcal{K}e^{2nz},$ $C_{tztz}=-(1/6)\mathcal{K},$ $C_{xyxy}%
=(1/6)\mathcal{K}e^{2nz}t^{2a_{1}},$ $C_{xzxz}=(1/6)\mathcal{K}t^{2a_{1}},$
and $C_{yzyz}=-(1/3)\mathcal{K}t^{2}e^{2nz},$ where $\mathcal{K=}\left(
n-1+a_{1}\right)  \left(  n+1-a_{1}\right)  .$ Therefore the Weyl invariant
yields, $I_{3}=\left(  4/3\right)  \mathcal{K}^{2}t^{-4}.$ The electric
invariant yields: $I_{4}=\left(  1/6\right)  \mathcal{K}^{2}t^{-4},$ while the
magnetic invariant, $I_{5},$ vanishes. The Weyl parameter yields,
$\mathcal{W}^{2}=\mathcal{K}^{2}/36(2+a_{1})^{4}=\mathrm{const.}$\textrm{,
(}note that $\mathcal{W}^{2}\rightarrow0$ is really small, so the model
isotropize) while the gravitational entropy yields, $P^{2}=\mathrm{const.}%
$\textrm{,} as it is expected\textrm{.}

\section{String cosmological model.}

The exponential expansion of the Universe (inflationary era) causes the
Universe to heat up to a very high temperature so the subsequent evolution of
the Universe is exactly as in hot BB model. Hence, the phase transition (as
the temperature falls below some critical temperature) in the early universe
causes topologically stable defects: vacuum domain walls, strings and
monopoles (see \cite{Z1} and \cite{v1}). But domain walls and monopoles are
disastrous for the cosmological models. Strings, on the other hand, causes no
harm, but can lead to very interesting astrophysical consequences (see
\cite{k}). Also the existence of a large scale network of strings the early
universe does not contradict the present-day observations. The vacuum strings
may generate density fluctuations sufficient to explain the galaxy formation
(see \cite{Z2}).

The relativistic treatment of strings was initiated by Letelier (see
\cite{L1}-\cite{L2}) and Stachel (see \cite{S}). Here we have considered
gravitational effects, arisen from strings by coupling of stress energy of
strings to the gravitational field. Letelier (see \cite{L2}) defined the
massive strings as the geometric strings (massless) with particles attached
along its expansions. Recently, Tikekar and Patel \cite{tk} have studied a
magnetic string Bianchi type $\mathrm{VI}$ model, Bali et al \cite{Bali} have
studied a magnetized bulk viscous massive string model and Pradhan and Bali
\cite{Prad} have worked with a magnetized bulk viscous string model with a
variable $\Lambda.$

The energy-momentum tensor, $T_{ij},$ for a cloud of massive strings is given
by%
\begin{equation}
T_{i}^{j}=\left(  \rho+p\right)  u^{j}u_{i}+pg_{i}^{j}-\lambda x^{j}x_{i},
\label{strin}%
\end{equation}
where $\rho(t)$ is the rest energy density, i.e. is the rest energy density of
the cloud of strings with particles attached to them ($p-$strings).
$\lambda(t)$ is the string tension density, which may be positive or negative,
$u^{i}$ is the four-velocity for the cloud particles. $x^{i}$ is the
four-vector which represents the strings direction which is the direction of
anisotropy and $\rho=\rho_{p}+\lambda,$ where $\rho_{p}$ denotes the particle
energy density (is the cloud rest energy density), i.e. the string tension
density is connected to the rest energy $\rho$ for a cloud of strings
($p-$strings) with particle attached to them by this relation (see \cite{p1}
and \cite{saha1}).

Since there is no direct evidence of strings in the present-day universe, we
are in general, interested in constructing models of a Universe that evolves
purely from the era dominated by either geometric string or massive strings
and ends up in a particle dominated era with or without remnants of strings.

Moreover the direction of strings satisfy the standard relations:%
\begin{align}
u^{j}u_{i}  &  =-x^{j}x_{i}=-1,\nonumber\\
u^{i}x_{i}  &  =0,\\
x^{i}  &  =\left(  0,a^{-1},0,0\right)  .\nonumber
\end{align}

In fact the choice of $x^{i}=\left(  0,0,0,d^{-1}(t)\right)  $ is the only one
possible since if we choice other direction then from the FE we get
$\lambda=0.$ It is customary to assume a relation between $\rho$ and $\lambda$
in accordance with the state equation for strings. The simplest one is a
proportionality relation $\rho=\alpha\lambda,$ where the most usual choices of
the constant $\alpha$ are the following ones (see \cite{saha1}). Geometric
strings. Nambu. $\alpha=1,\rho=\lambda,$ $\rho_{p}=0.$ $p-$strings or
Takabayasi strings%
\begin{equation}
\alpha=1+W, \label{es2}%
\end{equation}
with $W\geq0,$ iff $\rho_{p}=W\lambda.$ Reddy strings.$\alpha=-1.$ A more
general \textquotedblleft barotropic\textquotedblright\ equation of state is
$\rho=\rho\left(  \lambda\right)  ,\rho_{p}=\rho-\lambda.$ In the case of
Takabayasi strings if $W$ is very small, then only geometric strings appear.
On the other hand, if $W$ is infinitely large, then particles dominate over strings.

\bigskip

We shall consider the Takabayasi's equation of state, i.e. $\rho=\alpha
\lambda,$ with $\alpha=1+W,$ so the resulting FE are:\bigskip%
\begin{align}
\frac{a^{\prime}}{a}\frac{b^{\prime}}{b}+\frac{a^{\prime}}{a}\frac{d^{\prime}%
}{d}+\frac{d^{\prime}}{d}\frac{b^{\prime}}{b}+\frac{K}{d^{2}}  &  =8\pi
G\rho,\label{O1}\\
m\left(  \frac{a^{\prime}}{a}-\frac{d^{\prime}}{d}\right)  +n\left(
\frac{d^{\prime}}{d}-\frac{b^{\prime}}{b}\right)   &  =0,\label{O2}\\
\frac{b^{\prime\prime}}{b}+\frac{d^{\prime\prime}}{d}+\frac{d^{\prime}}%
{d}\frac{b^{\prime}}{b}-\frac{n^{2}}{d^{2}}  &  =-8\pi G\omega\rho
,\label{O3}\\
\frac{d^{\prime\prime}}{d}+\frac{a^{\prime\prime}}{a}+\frac{a^{\prime}}%
{a}\frac{d^{\prime}}{d}-\frac{m^{2}}{d^{2}}  &  =-8\pi G\omega\rho
,\label{O4}\\
\frac{b^{\prime\prime}}{b}+\frac{a^{\prime\prime}}{a}+\frac{a^{\prime}}%
{a}\frac{b^{\prime}}{b}+\frac{mn}{d^{2}}  &  =-8\pi G\left(  \omega
\rho-\lambda\right)  ,\label{O5}\\
\rho^{\prime}+\rho\left(  1+\omega\right)  H  &  =\lambda\frac{d^{\prime}}%
{d},\label{O6}\\
\lambda\left(  m-n\right)   &  =0. \label{O7}%
\end{align}

We have obtained eqs. (\ref{O6}-\ref{O7}) from the condition
$\operatorname{div}T=0.$ As is it observed eq. (\ref{O7}) bring us to simplify
the FE, since $m=n$, and then from eq. (\ref{O2}) we get a new constrain,
$a=b,$ so the FE collapse to the following ones:
\begin{align}
\left(  \frac{a^{\prime}}{a}\right)  ^{2}+2\frac{a^{\prime}}{a}\frac
{d^{\prime}}{d}-\frac{m^{2}}{d^{2}}  &  =8\pi G\rho,\label{nO1}\\
\frac{d^{\prime\prime}}{d}+\frac{a^{\prime\prime}}{a}+\frac{a^{\prime}}%
{a}\frac{d^{\prime}}{d}-\frac{m^{2}}{d^{2}}  &  =-8\pi G\omega\rho
,\label{nO2}\\
2\frac{a^{\prime\prime}}{a}+\left(  \frac{a^{\prime}}{a}\right)  ^{2}%
+\frac{m^{2}}{d^{2}}  &  =-8\pi G\left(  \omega\rho-\lambda\right)
,\label{nO3}\\
\rho^{\prime}+\rho\left(  1+\omega\right)  \left(  2\frac{a^{\prime}}{a}%
+\frac{d^{\prime}}{d}\right)   &  =\lambda\frac{d^{\prime}}{d}. \label{nO4}%
\end{align}

So, following the above procedure, and taking into account the SS solution,
given by eq. (\ref{restric}), we may find the following solution. From
(\ref{nO4}) we get:%
\[
\rho=\rho_{0}t^{-\beta},\qquad\beta=\left(  1+\omega\right)  \left(
a_{1}+a_{2}+1\right)  -\frac{1}{1+W}=2,
\]
and the algebraic system to solve is the following one:%
\begin{align}
a_{1}\left(  a_{1}-1\right)  +a_{1}-m^{2}  &  =-A\omega,\label{pal1}\\
2a_{1}\left(  a_{1}-1\right)  +a_{1}^{2}+m^{2}  &  =-A\left(  \omega-\frac
{1}{1+W}\right)  ,\label{pal2}\\
\left(  1+\omega\right)  \left(  2a_{1}+1\right)  -\frac{1}{1+W}  &  =2,
\label{pal3}%
\end{align}
with
\[
\rho_{0}=\frac{A}{8\pi G},\qquad A=a_{1}^{2}+2a_{1}-m^{2}.
\]

Therefore we may find the following solution:%
\begin{align}
a_{1}  &  =a_{2},\qquad m=\sqrt{\frac{a_{1}^{2}\left(  \omega+1\right)
+2a_{1}\omega}{\omega+1}}=n,\\
W  &  =-\frac{\omega\left(  2a_{1}+1\right)  +2\left(  a_{1}-1\right)
}{\omega\left(  2a_{1}+1\right)  +2a_{1}-1}.
\end{align}

Discussion: Since $W\geq0,$ and $a_{1}>0,$ $\left(  a_{1}^{2}\left(
\omega+1\right)  +2a_{1}\omega\geq0\right)  $ we get: \begin{widetext}%
\[%
\begin{array}
[c]{|l|l|l|}\hline
\omega & m\left(  \omega\right)  & W\left(  \omega\right) \\\hline\hline
1 & \sqrt{a_{1}^{2}+a_{1}}\geq0,\forall a_{1}\geq0 & -\frac{\left(
2a_{1}+1\right)  +2\left(  a_{1}-1\right)  }{\left(  2a_{1}+1\right)
+2a_{1}-1}\geq0,\forall a_{1}\in\left(  0,1/4\right)  \Longrightarrow
W\in\left(  0,\infty\right) \\\hline
1/3 & \sqrt{a_{1}^{2}+\frac{1}{2}a_{1}}\geq0,\forall a_{1}\geq0 &
-\frac{8a_{1}-5}{8a_{1}-2}\geq0,\forall a_{1}\in\left(  1/4,5/8\right)
\Longrightarrow W\in\left(  0,\infty\right) \\\hline
0 & a_{1} & -\frac{2\left(  a_{1}-1\right)  }{2a_{1}-1}\geq0,\forall a_{1}%
\in(1/2,1]\Longrightarrow W\in\left(  0,\infty\right) \\\hline
-1/3 & \sqrt{a_{1}\left(  a_{1}-1\right)  }\geq0,\forall a_{1}\geq1 &
-\frac{4a_{1}-7}{4a_{1}-4}\geq0,\forall a_{1}\in(1,7/4]\Longrightarrow
W\in\left(  0,\infty\right) \\\hline
-2/3 & \sqrt{a_{1}\left(  a_{1}-4\right)  }\geq0,\forall a_{1}\geq4 &
-\frac{2a_{1}-8}{2a_{1}-5}\geq0,\forall a_{1}\in(2.7,4]\Longrightarrow
\text{\textexclamdown !}\\\hline
\end{array}
\]
\end{widetext}

Therefore, by depending on the equation of state for the perfect fluid we have
a very different behaviour. The only nonphysical solution is $\omega=-2/3.$
The solution leads us to get the following metric:%
\begin{equation}
ds^{2}=-dt^{2}+t^{2a_{1}}e^{-2mz}dx^{2}+t^{2a_{1}}e^{2mz}dy^{2}+t^{2}dz^{2},
\label{B6s}%
\end{equation}
with%
\begin{equation}
\left(  1+\omega\right)  \left(  1+2a_{1}\right)  -\frac{1}{1+W}%
=\beta=2,\qquad\rho=\rho_{0}t^{-2}.
\end{equation}

In this case we have not obtained the corresponding Bianchi $\mathrm{III}$
solution since we get only a solution for $n=m\neq0.$ As it is observed the
metric (\ref{B6s}) has the same structure as the metric (\ref{B6c}), so it
will have the same curvature behavior.

Note that eq. (\ref{O7}) is quite strong since it says us that if we are
working with a Bianchi $\mathrm{III}$ metric, where $n=0,$ then the only one
(the unique) possibility is $m=0,$ in such a way that the metric collapses to
the Bianchi type $\mathrm{I}$. This means that there is not massive string
Bianchi $\mathrm{III}$ solution.

\section{String cosmological model with time varying \textquotedblleft
constants\textquotedblright.}

Following the above model we shall try to generalize it by considering the
possible time variation of the \textquotedblleft constants\textquotedblright%
\ $G$ and $\Lambda.$ In this case the FE\ are:%
\begin{align}
\frac{a^{\prime}}{a}\frac{b^{\prime}}{b}+\frac{a^{\prime}}{a}\frac{d^{\prime}%
}{d}+\frac{d^{\prime}}{d}\frac{b^{\prime}}{b}+\frac{K}{d^{2}}  &  =8\pi
G\rho+\Lambda,\label{h1}\\
m\left(  \frac{a^{\prime}}{a}-\frac{d^{\prime}}{d}\right)   &  =n\left(
\frac{b^{\prime}}{b}-\frac{d^{\prime}}{d}\right)  ,\label{h2}\\
\frac{b^{\prime\prime}}{b}+\frac{d^{\prime\prime}}{d}+\frac{d^{\prime}}%
{d}\frac{b^{\prime}}{b}-\frac{n^{2}}{d^{2}}  &  =-8\pi G\omega\rho
+\Lambda,\label{h3}\\
\frac{d^{\prime\prime}}{d}+\frac{a^{\prime\prime}}{a}+\frac{a^{\prime}}%
{a}\frac{d^{\prime}}{d}-\frac{m^{2}}{d^{2}}  &  =-8\pi G\omega\rho
+\Lambda,\label{h4}\\
\frac{b^{\prime\prime}}{b}+\frac{a^{\prime\prime}}{a}+\frac{a^{\prime}}%
{a}\frac{b^{\prime}}{b}+\frac{mn}{d^{2}}  &  =-8\pi G\left(  \omega
\rho-\lambda\right)  +\Lambda,\label{h5}\\
\rho^{\prime}+\rho\left(  1+\omega\right)  H  &  =\lambda\frac{d^{\prime}}%
{d},\label{h6}\\
\lambda\left(  m-n\right)   &  =0,\label{h7}\\
\Lambda^{\prime}  &  =-8\pi G^{\prime}\rho. \label{h8}%
\end{align}
where we have considered the condition $\operatorname{div}T=0,$ eqs.
(\ref{h6}-\ref{h7}), and therefore we have obtained eq. (\ref{h8}) as
additional restriction.

As above, we may simplify the FE, so they yield now as follows
\begin{align}
\left(  \frac{a^{\prime}}{a}\right)  ^{2}+2\frac{a^{\prime}}{a}\frac
{d^{\prime}}{d}-\frac{m^{2}}{d^{2}}  &  =8\pi G\rho+\Lambda,\label{iga01}\\
\frac{a^{\prime\prime}}{a}+\frac{d^{\prime\prime}}{d}+\frac{d^{\prime}}%
{d}\frac{a^{\prime}}{a}-\frac{m^{2}}{d^{2}}  &  =-8\pi G\omega\rho
+\Lambda,\label{iga02}\\
2\frac{a^{\prime\prime}}{a}+\left(  \frac{a^{\prime}}{a}\right)  ^{2}%
+\frac{m^{2}}{d^{2}}  &  =-8\pi G\left(  \omega\rho-\lambda\right)
+\Lambda,\label{iga03}\\
\rho^{\prime}+\rho\left(  1+\omega\right)  \left(  2\frac{a^{\prime}}{a}%
+\frac{d^{\prime}}{d}\right)   &  =\lambda\frac{d^{\prime}}{d},\label{iga04}\\
\Lambda^{\prime}  &  =-8\pi G^{\prime}\rho. \label{iga05}%
\end{align}
i.e. $m=n,$ and $a=b.$ By considering the SS solution given by eq.
(\ref{restric}) we find from eqs. (\ref{iga04}) and (\ref{es2})%
\begin{equation}
\rho=\rho_{0}t^{-\gamma},\quad\gamma=\left(  2a_{1}+1\right)  \left(
\omega+1\right)  -\frac{1}{1+W},
\end{equation}
where we shall assume that $\rho_{0}>0.$ From eq. (\ref{iga01}) we obtain:%
\begin{equation}
\Lambda=\frac{1}{c^{2}}\left[  At^{-2}-\frac{8\pi G}{c^{2}}\rho_{0}t^{-\gamma
}\right]  , \label{peta2}%
\end{equation}
where $A=a_{1}^{2}+2a_{1}-m^{2}.$

Now, taking into account eq. (\ref{iga05}) and eq. (\ref{peta2}), algebra
leads us to obtain%
\begin{equation}
G=G_{0}t^{\gamma-2},\qquad G_{0}=\frac{A}{4\pi\rho_{0}\gamma},
\end{equation}
and therefore the cosmological \textquotedblleft constant\textquotedblright%
\ behaves as:
\begin{equation}
\Lambda=A\left(  1-\frac{2}{\gamma}\right)  t^{-2}=\Lambda_{0}t^{-2}.
\end{equation}

With all these results, we find that the algebraic system to solve is the
following one:%
\begin{align}
a_{1}^{2}-m^{2}  &  =A\left(  1-\frac{2\left(  \omega+1\right)  }{\gamma
}\right)  ,\\
3a_{1}^{2}-2a_{1}+m^{2}  &  =A\left(  1-\frac{2\left(  \omega+1\right)
}{\gamma}-\frac{2}{\left(  1+W\right)  \gamma}\right)  ,
\end{align}
and where its solution is: $a_{1}=m=0,$ so we arrive to the conclusion that
there is not SS solution for this model.

\section{Conclusions.}

In this paper we have studied some Bianchi types $\mathrm{VI}_{0}%
\&\mathrm{III}$ models under the self-similarity hypothesis. We have started
by reviewing the \textquotedblleft classical\textquotedblright\ perfect fluid
solution already studied by Collins, Wainwright and Hsu. Nevertheless, we have
obtained two very restrictive solutions, which to the best of our knowledge
are new. The first of them corresponds to a Bianchi type $\mathrm{III}$ and is
only valid for $\omega\in\left(  -1/3,0\right)  $ in order to make $a_{2}>0.$
The second of them corresponds to Bianchi type $\mathrm{VI}$, and is only
valid for an equation of state $\omega\in\left(  -1/3,1\right)  $. Of course
we have also obtained the \textquotedblleft classical\textquotedblright%
\ solution, already studied by Collins and several authors. For all these
solutions we have studied their curvature behaviour, i.e. we have studied
their curvature invariants together with the Weyl tensor, the Weyl parameter,
$\mathcal{W},$ (showing that the models isotropize since $\left(
\mathcal{W}\rightarrow0\right)  )$ and we have also calculated the
gravitational entropy, $P^{2}$. In all the cases studied, $P^{2}%
=\mathrm{const}.,$ since we are working with a SS solution.

In the second of our studied models we have found two solutions for the vacuum
model. The first of the obtained solutions corresponds to Bianchi type
$\mathrm{VI,}$ and to the best of our knowledge it is new. By calculating its
curvature invariants, $\left(  I_{i}\right)  _{i=0}^{3},$ we have shown that
all of them vanish, nevertheless the Weyl tensor does not. The Weyl parameter
is really small and constant and the gravitational entropy, $P^{2}%
=\mathrm{\infty},$ since $I_{2}=0.$ This fact tells us that the definition of
the gravitational entropy does not work for vacuum models. The second of the
solutions obtained corresponds to Bianchi type $\mathrm{III}$ and it has
already been obtained by Hsu and Wainwright. It is a flat solution, for this
reason all its curvature invariants vanish.

With regard to a perfect fluid model with time-varying constants we have
obtained two solutions. Both solutions are valid for equation of state
$\omega\in(-1,1].$ In this case we have been able to enlarge the range of
validity for the equation of state. The first of the studied solution is of
Bianchi type $\mathrm{VI,}$ while the second one belongs to Bianchi type
$\mathrm{III.}$ In both solutions we have shown that if $G$ behaves as a
growing time function then $\Lambda$ is a \textquotedblleft\emph{positive}%
\textquotedblright\ decreasing time function. In the same way, if $G$ is
decreasing then $\Lambda$ behaves as a \textquotedblleft\emph{negative}%
\textquotedblright\ decreasing time function.

For the massive string cosmological model we have been able to obtain a
solution for Bianchi type $\mathrm{VI,}$ where the equations of state are
$\omega\in(-2/3,1]$ and $W\in\left(  0,\infty\right)  .$ As we have seen, it
is impossible to obtain a Bianchi type $\mathrm{III}$ solution. Such
restriction is obtained from the conservation equation, $\operatorname{div}%
T=0.$ In the last of our studied models, which is a massive string
cosmological model with time-varying constants, we have arrived at the
conclusion that there is no SS solution for this model, which is quite surprising.

\end{document}